\documentclass[10pt, conference]{IEEEtran}
\IEEEoverridecommandlockouts

\usepackage{cite}
\usepackage{amsmath,graphicx,amssymb,amsfonts}
\usepackage{amssymb}
\usepackage{bm}
\usepackage[inline]{enumitem}
\usepackage{algorithmic}
\usepackage{textcomp}
\usepackage{xcolor}
\usepackage[cmintegrals]{newtxmath}
\usepackage[nolist]{acronym}
\usepackage[hidelinks]{hyperref}
\usepackage{misc}
    
\hypersetup{
    colorlinks=true,
    linkcolor={link-blue},
    citecolor={link-blue},
    urlcolor={link-blue}
}
\graphicspath{{images/}}

\begin{acronym}[WSPSNR] 
    \acro{ad}[AD]{Arithmetic Decoder}
    \acro{ae}[AE]{Arithmetic Encoder}
    \acro{bpg}[BPG]{Better Portable Graphics}
    \acro{bpp}[bpp]{bits per pixel}
    \acro{bd}[BD]{Bj\o ntegaard-Delta}
    \acro{cm}[CM]{Context Model}
    \acro{cmp}[CMP]{Cube Map Projection}
    \acro{cnn}[CNN]{Convolutional Neural Network}
    \acro{conv}[Conv]{Convolution}
    \acro{erp}[ERP]{Equirectangular Projection}
    \acro{fp}[FP]{Factorized Prior}
    \acro{gdn}[GDN]{Generalized Divisive Normalization}
    \acro{healpix}[HEALPix]{Hierarchical Equal Area isoLatitude Pixelization}
    \acro{igdn}[IGDN]{Inverse Generalized Divisive Normalization}
    \acro{lrelu}[LReLU]{Leaky Rectified Linear Unit}
    \acro{mlp}[MLP]{Multilayer Perceptron}
    \acro{mse}[MSE]{Mean Squared Error}
    \acro{msh}[MSH]{Mean and Scale Hyperprior}
    \acro{nn}[NN]{Neural Network}
    \acro{oslo}[OSLO]{On-the-Sphere Learning for Omnidirectional Images}
    \acro{pa}[PA]{Parameter Aggregation}
    \acro{pmf}[PMF]{Probability Mass Function}
    \acro{psnr}[PSNR]{Peak Signal-to-Noise Ratio}
    \acro{q}[Q]{Quantization}
    \acro{rb}[RB]{Residual Block}
    \acro{rd}[RD]{Rate-Distortion}
    \acro{relu}[ReLU]{Rectified Linear Unit}
    \acro{wmse}[WMSE]{Weighted Mean Squared Error}
    \acro{sh}[SH]{Scale Hyperprior}
    \acro{spsnr}[S-PSNR]{Spherical Peak Signal-to-Noise Ratio}
    \acro{sgd}[SGD]{Stochastic Gradient Descent}
    \acro{tconv}[TConv]{Transposed Convolution}
    \acro{vae}[VAE]{Variational Autoencoder}
    \acro{vr}[VR]{Virtual Reality}
    \acro{vvc}[VVC]{Versatile Video Coding}
    \acro{wspsnr}[WS-PSNR]{Weighted-to-Spherically-Uniform Peak Signal-to-Noise Ratio}
\end{acronym}

\begin{document}

\title{OSLO-IC: On-the-Sphere Learned Omnidirectional Image Compression with Attention Modules and Spatial Context
\thanks{The authors gratefully acknowledge that this work has been supported by the Deutsche Forschungsgemeinschaft (DFG, German Research Foundation) under project number 418866191.}
}

\author{\IEEEauthorblockN{Paul~Wawerek-López\IEEEauthorrefmark{1}, Navid~Mahmoudian~Bidgoli\thanks{Navid M. Bidgoli is with Trimble Inc. This work was developed while he was with INRIA.}, Pascal~Frossard\IEEEauthorrefmark{2}, André~Kaup\IEEEauthorrefmark{1}, and Thomas~Maugey\IEEEauthorrefmark{3}} \IEEEauthorblockA{\IEEEauthorrefmark{1}Multimedia Communications and Signal Processing, Friedrich-Alexander-Universität Erlangen-Nürnberg, Germany} \IEEEauthorblockA{\IEEEauthorrefmark{2}École Polytechnique Fédérale de Lausanne (EPFL), Lausanne, Switzerland}
\IEEEauthorblockA{\IEEEauthorrefmark{3}Institut National de Recherche en Informatique et en Automatique (INRIA), Rennes, France}}

\maketitle

\begin{abstract}
Developing effective 360-degree (spherical) image compression techniques is crucial for technologies like virtual reality and automated driving.
This paper advances the state-of-the-art in on-the-sphere learning (OSLO) for omnidirectional image compression framework by proposing spherical attention modules, residual blocks, and a spatial autoregressive context model.
These improvements achieve a~$23.1\%$ bit rate reduction in terms of WS-PSNR BD rate.
Additionally, we introduce a spherical transposed convolution operator for upsampling, which reduces trainable parameters by a factor of four compared to the pixel shuffling used in the OSLO framework, while maintaining similar compression performance.
Therefore, in total, our proposed method offers significant rate savings with a smaller architecture and can be applied to any spherical convolutional application.
\end{abstract}


\section{Introduction}\label{sec:introduction}

The ongoing progress of technologies including \ac{vr} and autonomous driving results in an increasing amount of omnidirectional content.
Especially for consumer-oriented devices like \ac{vr} glasses, high resolution images are required to obtain a good visual impression at each viewing angle.
Thus, developing efficient compression techniques for omnidirectional images is necessary to store and transmit high-resolution omnidirectional images.
In the field of 2D~image and video compression, recent end-to-end learned methods based on autoencoders~\cite{balle2017factorizedprior, balle2018hyperprior, minnen2018, cheng2020, he2022elic, brand2024conditional-res} have led to promising compression performance.

However, a direct extension of \acp{cnn} to omnidirectional images is not straightforward since various representations exist.
One method consists in mapping the sphere to one or multiple planes as a preprocessing step.
The most popular among these mappings is the \ac{erp}.
Due to the inevitable geometrical distortions introduced by such mappings, the interpixel correlations become position-dependent.
To deal with the distortions of the \ac{erp}, the authors of~\cite{li2022} introduced a latitude-dependent loss function as well as an additional latitude adaptive scale network.
For the same reason, using multiple motion models for video compression was investigated in~\cite{regensky2023multimotion}.
In contrast, the authors of~\cite{bidgoli2022oslo} defined convolution operations directly on a spherical representation based on \acs{healpix} sampling~\cite{gorski2005healpix}.
This framework is called \ac{oslo}, which enhances the performance of \acp{cnn} for omnidirectional image compression and denoising compared to training the equivalent 2D~models on \ac{erp} images~\cite{bidgoli2022oslo,fermanian2023sdrunet}.
However, recent advancements in 2D image compression models, including attention modules, residual blocks as nonlinearities, and a spatial context model for on-the-sphere learned image compression, have not yet been integrated into the OSLO framework.
Additionally, \ac{oslo} only supports periodic pixel shuffling as an unpooling operation, resulting in four times the number of parameters compared to other well-known unpooling methods.
Defining computationally efficient on-the-sphere operations is challenging since the \acs{healpix} images are stored in vectors, containing only information about the direct neighborhoods of each pixel.

In this paper, we build on \ac{oslo} and propose an updated spherical end-to-end learned image compression model containing attention modules~\cite{cheng2020}, the replacement of \ac{gdn} layers~\cite{balle2016gdn} by residual blocks~\cite{he2022elic}, and a spatial autoregressive context model~\cite{minnen2018}.
The context model is realized by defining a spherical masked convolution.
Compared to the spherical scale hyperprior model, the proposed model saves~$23\%$ of the bit rate in terms of \acs{bd} rate~\cite{bjontegaard2001bdrate,herglotz2024bjontegaardbible} calculated on \acs{wspsnr}~\cite{sun2017wspsnr}.
In addition, we define a spherical transposed convolution, reducing the number of parameters by a factor of four while maintaining its expressiveness.
Since transposed convolutions are often used for other tasks including image generation or denoising, our proposed spherical transposed convolution reduces the complexity of many other spherical models.\footnote{The source code is available at \textit{\href{https://gitlab.inria.fr/on-the-sphere-learning-for-omnidirectional-images-oslo/oslo-360-degree-image-compression}{www.gitlab.inria.fr/on-the-sphere-learning-for-omnidirectional-images-oslo/oslo-360-degree-image-compression}}}

\section{On-the-Sphere learned Image Compression}\label{sec:groundwork}

\subsection{HEALPix Sampling}\label{subsec:healpix}
Using the \ac{healpix} proposed in~\cite{gorski2005healpix}, each pixel covers the same area on the sphere. Initially, the sphere is divided into $12$ equal-area base tiles (red lines in~\cref{fig:sphere}). Higher resolutions are achieved by recursively splitting each pixel into $4$ sub-pixels of equal area. The resolution parameter~$N_\mathrm{side}$ represents the number of pixels per base tile, resulting in a total of $N_\mathrm{pix}=12\cdot N_\mathrm{side}^2$ pixels. Pixel borders for $N_\mathrm{side} \leq 8$ are shown in~\cref{fig:sphere}. Each pixel has $8$ neighbors, except for $24$ pixels with $7$ neighbors located at the intersections of polar and equatorial tiles. This number of exceptional pixels is constant, making their effect negligible at high resolutions. Pixel indices follow a \textit{nested} tree structure, where a parent pixel index $[\dots p_2p_1p_0]_2$ (in binary notation) generates child pixel indices by appending two bits~$b_1b_0$, resulting in $[\dots p_2p_1p_0b_1b_0]_2$.

\subsection{Spherical Convolution}\label{subsec:spherical-conv}
\begin{figure}[!t]
    \centering
    \includegraphics[width=.5\linewidth]{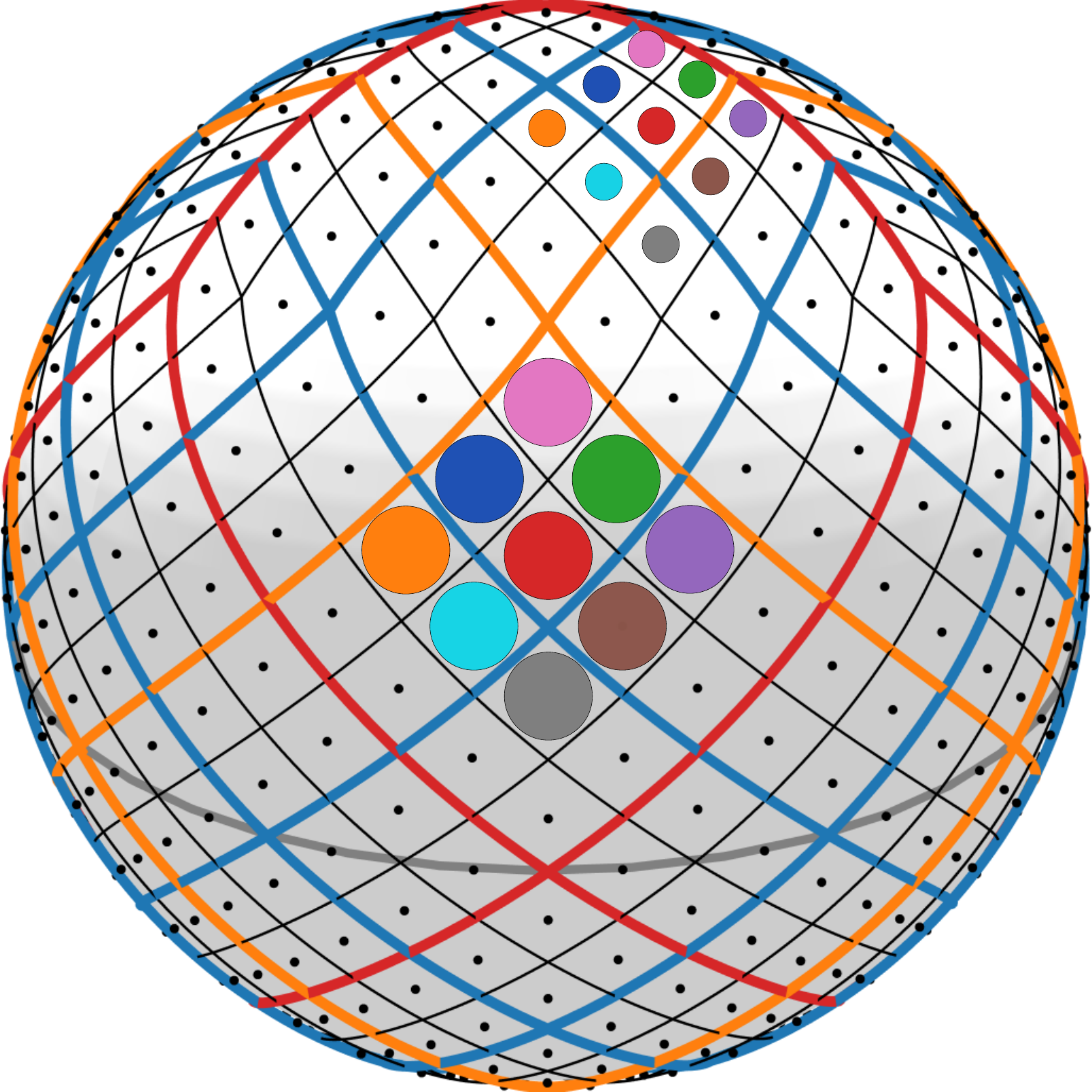}
    \caption{\acs{healpix} sampled sphere with resolution~$N_\mathrm{side}=8$. Red, orange and blue lines show the pixel boundaries for $N_\mathrm{side}=1, 2,4$, respectively. The gray line indicates the equator. Additionally, a spherical convolution kernel is shown at two positions on the sphere.}
    \label{fig:sphere}
\end{figure}

The \ac{oslo} framework proposed in~\cite{bidgoli2022oslo} defines spherical convolution as follows:
Let~$L_\mathrm{in}$ and~$L_\mathrm{out}$ be the number of input and output features.
The input features at pixel index $i=1,\dots,N_\mathrm{pix}$ are $\bm{x}_i\in\mathbb{R}^{L_\mathrm{in}}$, and the weights of the $l$-th kernel for the $k$-th neighbor are $\bm{\Theta}_k^l\in\mathbb{R}^{L_\mathrm{in}}$, with $k=1,\dots,8$ and $l=1,\dots,L_\mathrm{out}$.
The kernel shape is shown in~\cref{fig:sphere}, with each color representing a different kernel entry.
The output feature~$l$ of the spherical convolution is~\cite{bidgoli2022oslo}:
\begin{equation}\label{eq:sdpa-conv}
    x_i^l = \langle\bm{\Theta}_0^l,\bm{x}_i\rangle + \sum_{k=1}^8 \langle\bm{\Theta}_k^l,\bm{x}_{\mathcal{N}_i(k)}\rangle \cdot w_{\mathcal{N}_i(k),i},
\end{equation}
where~$\mathcal{N}_i(k)$ is the index of the $k$-th neighbor of pixel~$i$ and~$w_{\mathcal{N}_i(k),i}$ is~$0$ if~$\mathcal{N}_i(k)$ is missing, otherwise~$1$. Only $1$-hop convolutions are supported due to computational complexity and storage demands, but larger receptive fields can be achieved by performing $n$ subsequent $1$-hop convolutions and summing the outputs.

This spherical convolution meets three key requirements~\cite{bidgoli2022oslo}: 1)~\textit{Consistency}: The filter orientation towards the north pole, interpixel correlations, and pixel areas remain nearly constant over the sphere in \ac{healpix} sampling. 2)~\textit{Expressiveness}: The anisotropic filter makes distinction between neighboring pixels, similar to standard 2D convolution. 3)~\textit{Efficiency}: The computation scales linearly with the number of pixels.

\subsection{End-to-End Learned Image Compression}\label{subsec:e2e-learned-image-compression}
Most recent end-to-end learned image compression models are based on the scale hyperprior model introduced in~\cite{balle2018hyperprior}.
This model consists of two autoencoders: the image encoder and decoder~$(e,d)$, and the statistics encoder and decoder~$(e_\mathrm{s},d_\mathrm{s})$, known as the hyperprior network.

The image encoder~$e$ generates a latent space~$\bm{y}$ from the input~$\bm{x}$.
The latent space is quantized through rounding, but since rounding is not differentiable, it is emulated by additive uniform noise \textit{during training}~\cite{balle2016}.
The hyperprior network estimates the likelihood of~$\bm{y}$ using a Gaussian entropy model to improve the performance of the lossless arithmetic encoder and decoder. 
More precisely, the latent space~$\bm{y}$ is fed into~$e_\mathrm{s}$ to produce a representation~$\bm\nu$ of the distribution of standard deviations~$\sigma$.
This representation~$\bm\nu$ is quantized to~$\hat{\bm\nu}$ and losslessly compressed by an arithmetic encoder, forming the hyperprior bitstream transmitted as side information.
The scale estimations~$\hat{\sigma}$ are obtained by decoding the hyperprior bitstream with an arithmetic decoder and $d_\mathrm{s}$.
The arithmetic encoder and decoder use a Gaussian entropy model based on~$\hat{\sigma}$ to generate the image bitstream and reconstruct the quantized latent space~$\hat{\bm{y}}$, respectively.
The reconstructed image~$\hat{\bm{x}}$ is obtained by feeding~$\hat{\bm{y}}$ into~$d$.

During training, the model optimizes the trade-off between the bit rate~$R$ and the distortion~$D$ by minimizing the loss function: 
\begin{equation}\label{eq:loss} 
\mathcal{L} = R(\hat{\bm{y}},\hat{\bm\nu})+\lambda D(\bm{x},\hat{\bm{x}}), 
\end{equation}
where~$\lambda$ denotes the quality parameter. A higher~$\lambda$ places more importance on reconstruction quality. The distortion~$D$ is calculated from the mean squared error between the input~$\bm{x}$ and its reconstruction~$\hat{\bm{x}}$.

Various modifications of the scale hyperprior model have been proposed to improve compression performance.
Minnen~\textit{et al.} introduced a spatial autoregressive context model in~\cite{minnen2018}, estimating mean values in addition to standard deviations.
This spatial context model uses already decoded parts of the quantized latent space $\hat{\bm{y}}$ to predict future mean and standard deviation values, implemented with a masked convolution kernel and a parameter aggregation network.
Sequential decoding is required due to causality constraints in the spatial context model.
Channel-wise context models for faster parallel decoding were investigated in~\cite{Minnen2020channel-cm,Liang2021channel-grouped-cm,he2022elic}, but are not the focus here.
In~\cite{cheng2020}, Cheng~\textit{et al.} included attention modules in the image autoencoder, reducing bits for simple regions and spending more bits on challenging parts~\cite{cheng2020}.
He~\textit{et al.} proposed replacing (inverse) \ac{gdn} layers~\cite{balle2016gdn} with three consecutive residual blocks, which are more effective for Gaussian density modeling due to larger receptive fields~\cite{he2022elic}.

\begin{figure*}[!t]
    \centering
    \def\svgwidth{\linewidth}
    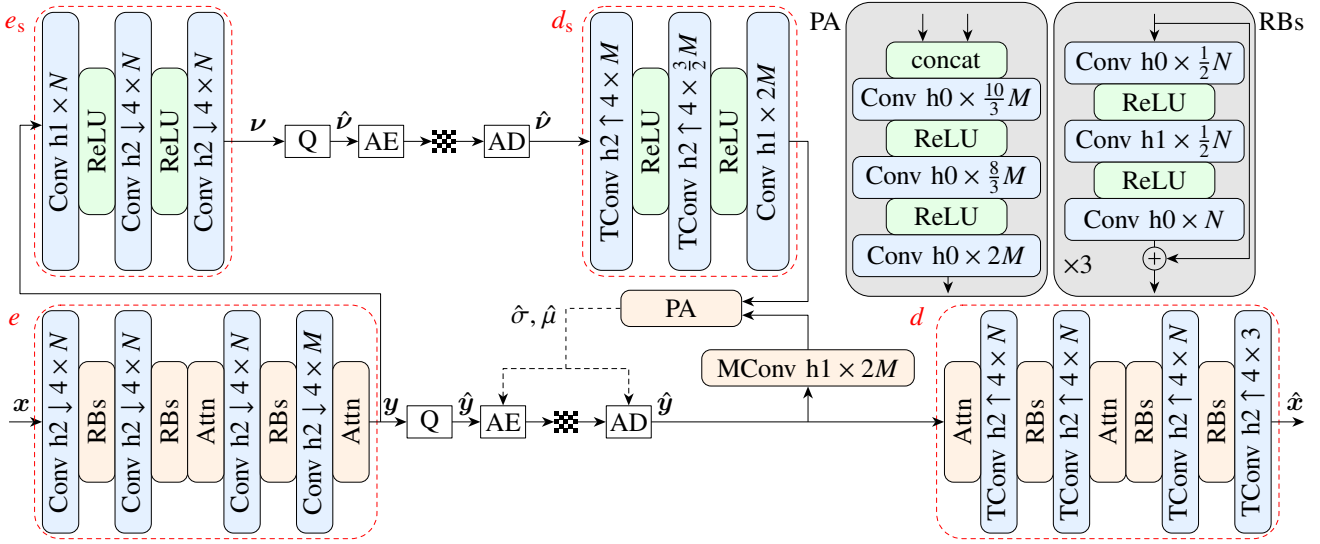
    \caption{Extended spherical model with Parameter Aggregation (PA) and Residual Blocks (RBs). Attention modules (Attn) follow the simplified structure in~\cite{cheng2020}. The description "Conv h$n\downarrow 4\times N$" represents a spherical $n$-hop convolution downsampled by~$4$ with $N$ filters. "TConv h$n\uparrow 4\times N$" represents a spherical $n$-hop transposed convolution upsampled by~$4$ with $N$ filters.}
    \label{fig:model}
\end{figure*}

\section{Extended Model}\label{sec:proposal}

In the following, we build our extended spherical compression model as shown in \cref{fig:model}, where $0$-, $1$-, and $2$-hop convolutions are the equivalents of $1\times1$, $3\times3$, and $5\times5$ convolutions for 2D images.

The spherical attention module and residual blocks are only built from $0$- and $1$-hop convolutions as well as element-wise operations.
For reference, we use the simplified attention module from~\cite{cheng2020}.
Let $\bm{x}$ denote the input of the spherical attention layer, then the output can be calculated as $\bm{x}+f^1_\mathrm{RBs}(\bm{x})\odot \sigma(f_{\mathrm{h}0}(f^2_\mathrm{RBs}(\bm{x})))$, where $f^1_\mathrm{RBs}(\cdot)$ and $f^2_\mathrm{RBs}(\cdot)$ denote two different stacks of three residual blocks as indicated in the top-right (\acsp{rb}) of~\cref{fig:model}, $f_{\mathrm{h}0}(\cdot)$ denotes a $0$-hop convolution, and $\odot$ denotes an element-wise multiplication.
The calculation of $0$-hop convolutions is a special case of~\cref{eq:sdpa-conv} when only the center of the kernel $\bm{\Theta}_0^l$ is considered.
The computation of $0$-hop convolutions thus includes only a single matrix multiplication.
Therefore, all necessary operations are defined inside the previous \ac{oslo} framework.
The resulting structure of the image autoencoder is adopted from~\cite{he2022elic}.

\begin{figure}[!t]
    \centering
    \def\svgwidth{.85\linewidth}
    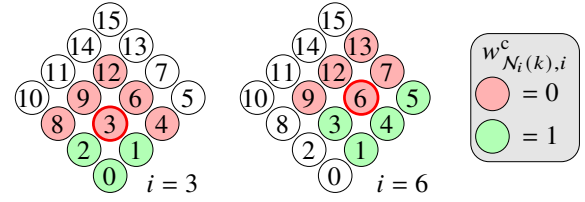
    \caption{Proposed causal $1$-hop convolution mask $w_{\mathcal{N}_i(k),i}^\mathrm{c}$ for the spatial context model at two different pixel positions~$i$. The pixel indices follow the nested \ac{healpix} indexing scheme. For example, when computing the convolution for pixel~$i=6$, the neighboring pixels $i= 1, 3, 4, 5$ have already been processed and are included in the masked convolution.}
    \label{fig:mconv}
\end{figure}

For the spatial context model, we define the spherical masked convolution by replacing the mask~$w_{\mathcal{N}_i(k),i}$ in~\cref{eq:sdpa-conv} by~$w^\prime_{i,k} = w_{\mathcal{N}_i(k),i}\cdot w^\mathrm{c}_{\mathcal{N}_i(k),i}$ with the position-dependent causal mask:
\begin{equation}
    w^\mathrm{c}_{\mathcal{N}_i(k),i} = \begin{cases}
        0, & \mathcal{N}_i(k) \geq i\\
        1, & i<\mathcal{N}_i(k)
    \end{cases},
\end{equation}
which is illustrated for a masked $1$-hop convolution at two different positions on the sphere in~\cref{fig:mconv}.
We also change the spherical scale hyperprior to also estimate the distribution of mean values by adapting the output features of the convolutions inside the decoder of statistics~$d_\mathrm{s}$.
The parameter aggregation module, which combines the output of~$d_\mathrm{s}$ with the context model for the prediction of mean and standard deviation distributions, consists only of $0$-hop convolutions and can thus be implemented with the previous \ac{oslo} framework.

\begin{figure}[!t]
    \centering
    \def\svgwidth{\linewidth}
    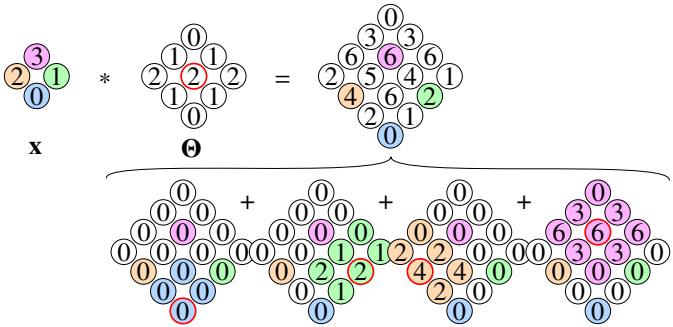
    \caption{Proposed spherical transposed convolution operation with input~$\mathbf{x}$ and filter kernel~$\mathbf{\Theta}$. The kernel gets multiplied by each input pixel value and added to the high-resolution output at the corresponding pixel position.}
    \label{fig:tconv}
\end{figure}

Additionally, we modify the unpooling operation by replacing the pixel shuffling operation with a spherical transposed convolution, illustrated in \cref{fig:tconv}.
The idea is to multiply the weights~$\mathbf{\Theta}$ with the input values~$\mathbf{x}_i$ for each pixel~$i$.
Each multiplied kernel is then added to the high-resolution output with the kernel center placed at the child pixels with index~$i^\prime=[\dots i_2i_1i_000]_2$ in binary notation.
To make this operation computationally efficient, we calculate matrix multiplications of the input resolution signals as~$\mathbf{x}\mathbf{\Theta}_0$ for the kernel center and~$\mathbf{x}\mathbf{\Theta}_k$ for each neighbor with~$\mathbf{x}\in\mathbb{R}^{N_\mathrm{pix}\times L_\mathrm{in}}$ and~$\mathbf{\Theta}_0,\mathbf{\Theta}_k\in\mathbb{R}^{L_\mathrm{in}\times L_\mathrm{out}}$.
The resulting~$N_\mathrm{pix}\times L_\mathrm{out}$ values are added to the high-resolution output image at the positions~$i^\prime$ for the center and~$\mathcal{N}_{i^\prime}(k)$ for each neighbor~$k$.

The proposed transposed convolution kernel thus consists of $(9L_\mathrm{in}+1)L_\mathrm{out}$ parameters with additional bias.
In contrast, using pixel shuffling as unpooling operation requires~$4\cdot L_\mathrm{out}$ output channels to provide the same output resolution, and therefore four times the number of parameters, due to the rearrangement of the output channels after the convolution.

\section{Performance Evaluation}\label{sec:performance}

To evaluate our extended spherical model, we use the same train and test setup as the authors of~\cite{bidgoli2022oslo}.
We use~$2170$ \ac{erp} images of resolution~$9104\times4552$ from the SUN360 dataset~\cite{xiao2012sun360} and split them into~$1737$ train, $10$ validation, and~$423$ test images with the identical split used in~\cite{bidgoli2022oslo}.
We resample the \ac{erp} images to \ac{healpix} with resolution~$N_\mathrm{side}=2^{10}$.
During training, random patches equivalent to a size of~$256\times256$ are fed into the model in batches of size~$10$.
We train our models for~$1000$ epochs with a fixed learning rate of~$10^{-4}$ during the first~$800$ epochs and reduce the learning rate based on the validation loss during the last 200 epochs.

\begin{figure}[!t]
    \centering
\begin{tikzpicture}

\definecolor{darkgray176}{RGB}{176,176,176}
\definecolor{green}{RGB}{0,128,0}
\definecolor{lightgray204}{RGB}{204,204,204}

\begin{axis}[
height=1.2\linewidth,
legend cell align={left},
legend style={
  fill opacity=0.8,
  draw opacity=1,
  text opacity=1,
  at={(1,0)},
  anchor=south east,
  draw=lightgray204
},
tick pos=left,
width=\linewidth,
x grid style={darkgray176},
xlabel={$R$ [bpp]},
xmajorgrids,
xmin=0, xmax=1.35,
xtick style={color=black},
xtick distance=.2,
ytick distance=2,
y grid style={darkgray176},
ylabel={\acs{wspsnr} [dB]},
ymajorgrids,
ymin=23, ymax=42, 
ytick style={color=black},
]
\addplot+ [semithick, blue, scatter=true, dashed, mark=o, mark size=2.555, mark options={solid},
point meta=explicit symbolic,
scatter/@pre marker code/.style={/tikz/mark size=\pgfplotspointmeta},
    scatter/@post marker code/.style={}]
table [meta index=2]{%
0.0335519003868103 25.25 2.169
0.06120943625768025 27.17 2.169
0.10166162808736166 29.05 2.169
0.1584190503756205 30.79 2.169
0.23666004260381066 32.4 2.169
0.34797880093256633 34.22 2.169
0.4950086386998494 36.02 2.169
0.6946404639879863 37.82 2.555
0.9605764293670654 39.6 2.555
1.248831938902537 40.85 2.555
};
\addlegendentry{\acs{sh} ($0.0\%$, $22.3\mathrm M$)}
\addplot+ [semithick, lightblue, dotted, scatter=true, mark=o, mark size=1.197, mark options={solid},
point meta=explicit symbolic,
scatter/@pre marker code/.style={/tikz/mark size=\pgfplotspointmeta},
    scatter/@post marker code/.style={}]
table [meta index=2]{%
0.04464107275009155 26.03 1.080
0.10899439573287963 29.17 1.080
0.24696368137995403 32.6 1.080
0.3553731028238933 34.27 1.080
0.5118899512290954 36.06 1.080
0.7194700551033021 37.95 1.197
0.9816564949353536 39.85 1.197
1.3204380695025126 41.52 1.197
};
\addlegendentry{\acs{sh}\_\acs{tconv} ($1.3\%$, $5.4\mathrm M$)}
\addplot+ [semithick, green, scatter=true, mark=o, mark size=2.998, mark options={solid},
point meta=explicit symbolic,
scatter/@pre marker code/.style={/tikz/mark size=\pgfplotspointmeta},
    scatter/@post marker code/.style={}]
table [meta index=2]{%
0.03160603205362956 25.51 2.443
0.09923899173736572 29.35 2.443
0.22570455074310303 32.6 2.443
0.3250076675415039 34.4 2.443
0.4496451783180237 36.05 2.443
0.626808557510376 37.76 2.998
0.861548048655192 39.5 2.998
1.1446627593040466 40.94 2.998
};
\addlegendentry{\acs{sh}\_Attn\_RB ($-9.9\%$, $27.8\mathrm M$)}
\addplot+ [semithick, lightgreen, dotted, scatter=true, mark=o, mark size=1.640, mark options={solid},
point meta=explicit symbolic,
scatter/@pre marker code/.style={/tikz/mark size=\pgfplotspointmeta},
    scatter/@post marker code/.style={}]
table [meta index=2]{%
0.03739121754964193 25.7 1.354
0.10033527374267577 29.24 1.354
0.2300092911720276 32.77 1.354
0.33106895446777346 34.48 1.354
0.4630992817878723 36.17 1.354
0.6426794656117757 38.0 1.640
0.8704190142949422 39.76 1.640
1.208016860485077 41.44 1.640
};
\addlegendentry{\acs{sh}\_Attn\_RB\_TConv ($-8.9\%$, $10.9\mathrm M$)}
\addplot+ [semithick, red, scatter=true, mark=o, mark size=5.992, mark options={solid},
point meta=explicit symbolic,
scatter/@pre marker code/.style={/tikz/mark size=\pgfplotspointmeta},
    scatter/@post marker code/.style={}]
table [meta index=2]{%
0.02603946050008138 26.21 3.789
0.08912585973739624 29.87 3.789
0.209273738861084 33.01 3.789
0.30279115358988445 34.54 3.789
0.42773613691329954 36.12 3.789
0.6014846245447795 37.79 5.992
0.8226803477605183 39.33 5.992
1.1172136227289835 41.15 5.992
};
\addlegendentry{Proposed ($-23.1\%$, $64.9\mathrm M$)}
\addplot+ [semithick, lightred, dotted, scatter=true, mark=o, mark size=2.26, mark options={solid},
point meta=explicit symbolic,
scatter/@pre marker code/.style={/tikz/mark size=\pgfplotspointmeta},
    scatter/@post marker code/.style={}]
table [meta index=2]{%
0.03291043043136597 26.66 1.649
0.08954503695170085 29.87 1.649
0.2124264701207479 33.02 1.649
0.30840940634409586 34.68 1.649
0.436442875067393 36.29 1.649
0.6062089633941651 38.01 2.260
0.8366147788365682 39.84 2.260
1.134238433043162 41.47 2.260
};
\addlegendentry{Proposed\_TConv ($-23.0\%$, $18.6\mathrm M$)}
\end{axis}

\end{tikzpicture}
    \caption{RD curves of different spherical models. The first number in parentheses is the \acs{bd}-rate compared to the spherical Scale Hyperprior (\acs{sh}) model (negative values show an improvement).
    Marker sizes scale with the number of trainable parameters (second number in parentheses).}
    \label{fig:rd-plot}
\end{figure}
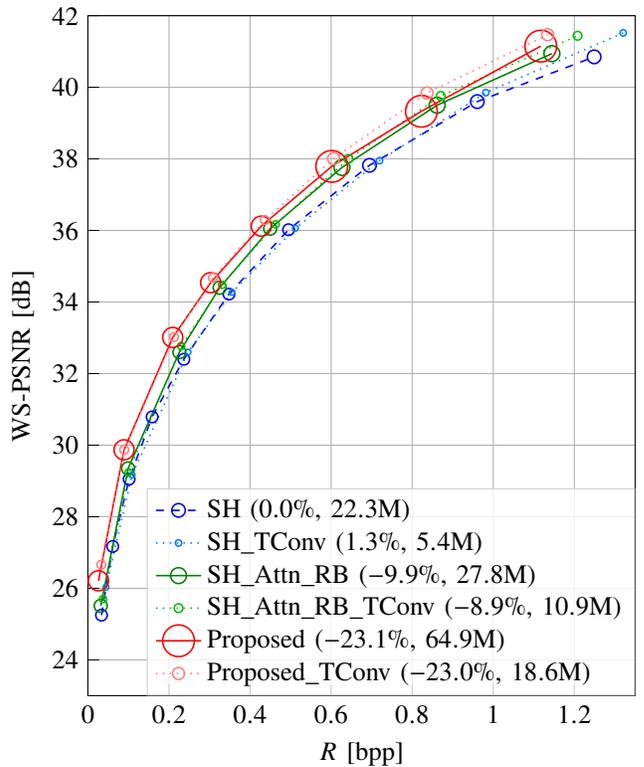

As a quantitative measurement of the reconstruction quality, we use the \ac{wspsnr}~\cite{sun2017wspsnr}.
The \ac{wspsnr} places more importance to pixels with large area and, therefore, is stronger correlated with subjective quality than \acsu{psnr}. Similar results were obtained for Spherical \acs{psnr} (\acs{spsnr})~\cite{yu2015spsnr}, which is not discussed here due to space constraints.
To evaluate the compression performance, we plot the \ac{rd} curves of each model in~\cref{fig:rd-plot}, where the \ac{wspsnr} values for different quality parameters $\lambda$ are averaged over the test images and plotted over the rate, given in \ac{bpp}.\footnote{$\lambda\in\{0.0005,0.0018,0.0067,0.0130,0.025,0.0483,0.0932,0.18\}$ and optionally $\lambda\in\{0.0009,0.0035\}$.}

\cref{fig:rd-plot} also contains the \ac{bd} rates~\cite{bjontegaard2001bdrate,herglotz2024bjontegaardbible} in $\%$ and model sizes in number of trainable parameters, which are given for the three highest quality parameters $\lambda$ for each model.\footnote{$(N,M)=(142,270)$ for $\lambda\geq 0.0483$, and otherwise $(128,192)$.}
We can see in the red curve that our model including all extensions outperforms the spherical scale hyperprior model from~\cite{bidgoli2022oslo} at all quality levels and saves~$23.1\%$ of the bit rate in terms of \ac{bd} rate.
However, the number of trainable parameters is considerably higher, which is mainly caused by the pixel shuffling operation.
Using our transposed convolution instead (dotted red curve), reduces the model size by more than a factor of~$3$ while providing almost the same rate saving of~$23.0\%$ in terms of \ac{bd} rate.
We can also see from the green curves that the spherical attention modules and residual blocks reduce the bit rate by almost~$10\%$ while only leading to a moderate increase of the trainable parameters of roughly~$25\%$.
Similarly to the full model, our transposed convolution (dotted green curve) reduces the model size by a factor of~$2.6$ while only leading to a minor decrease in \ac{bd} rate.
Using transposed convolution for the original spherical scale hyperprior model (dotted blue curve) reduces the model size by a factor of around~$4.1$ with~$1.3\%$ more bits needed.
Despite the transposed convolution increasing the bit rate by a small portion, all models using spherical transposed convolution achieve better reconstruction quality at higher bit rates compared to the equivalent models using pixel shuffling.

\section{Conclusion}\label{sec:conclusion}

In this paper, we showed that recent 2D image compression models significantly improve on-the-sphere learned compression models based on the \ac{oslo} framework.
By incorporating attention modules, residual blocks as nonlinearities, and a spatial context model into the previous spherical scale hyperprior model, we save~$23.1\%$ of the bit rate based on \ac{wspsnr}.
For the unpooling operation, the proposed transposed convolution reduces the required number of parameters by a factor of~$4$ while maintaining almost the same filter expressiveness.
In the future, we aim to extend the \ac{oslo} framework by a computationally efficient channel-wise context model as well as different model architectures including transformer-based networks.

\bibliographystyle{IEEEtran}
\bibliography{src}

\begin{thebibliography}{10}
\providecommand{\url}[1]{#1}
\csname url@samestyle\endcsname
\providecommand{\newblock}{\relax}
\providecommand{\bibinfo}[2]{#2}
\providecommand{\BIBentrySTDinterwordspacing}{\spaceskip=0pt\relax}
\providecommand{\BIBentryALTinterwordstretchfactor}{4}
\providecommand{\BIBentryALTinterwordspacing}{\spaceskip=\fontdimen2\font plus
\BIBentryALTinterwordstretchfactor\fontdimen3\font minus \fontdimen4\font\relax}
\providecommand{\BIBforeignlanguage}[2]{{%
\expandafter\ifx\csname l@#1\endcsname\relax
\typeout{** WARNING: IEEEtran.bst: No hyphenation pattern has been}%
\typeout{** loaded for the language `#1'. Using the pattern for}%
\typeout{** the default language instead.}%
\else
\language=\csname l@#1\endcsname
\fi
#2}}
\providecommand{\BIBdecl}{\relax}
\BIBdecl

\bibitem{balle2017factorizedprior}
J.~Ballé, V.~Laparra, and E.~P. Simoncelli, ``End-to-end optimized image compression,'' in \emph{Proc. International Conference on Learning Representations (ICLR)}, 2017.

\bibitem{balle2018hyperprior}
J.~Ball\'e, D.~Minnen, S.~Singh, S.~J. Hwang, and N.~Johnston, ``Variational image compression with a scale hyperprior,'' in \emph{Proc. International Conference on Learning Representations (ICLR)}, 2018.

\bibitem{minnen2018}
D.~Minnen, J.~Ball\'e, and G.~D. Toderici, ``Joint autoregressive and hierarchical priors for learned image compression,'' in \emph{Proc. Advances in Neural Information Processing Systems (NeurIPS)}, vol.~31, 2018, pp. 10\,771--10\,780.

\bibitem{cheng2020}
Z.~Cheng, H.~Sun, M.~Takeuchi, and J.~Katto, ``Learned image compression with discretized gaussian mixture likelihoods and attention modules,'' in \emph{Proc. IEEE/CVF Conference on Computer Vision and Pattern Recognition (CVPR)}, 2020.

\bibitem{he2022elic}
D.~He, Z.~Yang, W.~Peng, R.~Ma, H.~Qin, and Y.~Wang, ``Elic: Efficient learned image compression with unevenly grouped space-channel contextual adaptive coding,'' in \emph{Proc. IEEE/CVF Conference on Computer Vision and Pattern Recognition (CVPR)}, 2022.

\bibitem{brand2024conditional-res}
F.~Brand, J.~Seiler, and A.~Kaup, ``Conditional residual coding: A remedy for bottleneck problems in conditional inter frame coding,'' \emph{IEEE Transactions on Circuits and Systems for Video Technology (CSVT)}, vol.~34, no.~7, pp. 6445--6459, 2024.

\bibitem{li2022}
M.~Li, J.~Li, S.~Gu, F.~Wu, and D.~Zhang, ``End-to-end optimized 360° image compression,'' \emph{IEEE Transactions on Image Processing}, vol.~31, pp. 6267--6281, 2022.

\bibitem{regensky2023multimotion}
A.~Regensky, C.~Herglotz, and A.~Kaup, ``Multi-model motion prediction for 360-degree video compression,'' \emph{IEEE Access}, vol.~11, pp. 117\,004--117\,017, 2023.

\bibitem{bidgoli2022oslo}
N.~Mahmoudian~Bidgoli, R.~G. de~A.~Azevedo, T.~Maugey, A.~Roumy, and P.~Frossard, ``Oslo: On-the-sphere learning for omnidirectional images and its application to 360-degree image compression,'' \emph{IEEE Transactions on Image Processing}, vol.~31, pp. 5813--5827, 2022.

\bibitem{gorski2005healpix}
K.~M. Gorski, E.~Hivon, A.~J. Banday, B.~D. Wandelt, F.~K. Hansen, M.~Reinecke, and M.~Bartelmann, ``Healpix: A framework for high-resolution discretization and fast analysis of data distributed on the sphere,'' \emph{The Astrophysical Journal}, vol. 622, no.~2, p. 759, 2005.

\bibitem{fermanian2023sdrunet}
R.~Fermanian, T.~Maugey, and C.~Guillemot, ``Spheredrunet: A spherical denoiser for omnidirectional images,'' in \emph{Proc. IEEE International Symposium on Mixed and Augmented Reality Adjunct (ISMAR-Adjunct)}, 2023, pp. 417--422.

\bibitem{balle2016gdn}
J.~Ball\'e, V.~Laparra, and E.~P. Simoncelli, ``Density modeling of images using a generalized normalization transformation,'' in \emph{Proc. International Conference on Learning Representations (ICLR)}, 2016.

\bibitem{bjontegaard2001bdrate}
G.~Bj{\o}ntegaard, ``Calculation of average psnr differences between rd-curves, {{VCEG-M33}},'' in \emph{Proc. 13th {{Meet}}. {{Video Coding Experts Group}}}, Mar. 2001, pp. 1--5.

\bibitem{herglotz2024bjontegaardbible}
C.~Herglotz, H.~Och, A.~Meyer, G.~Ramasubbu, L.~Eichermüller, M.~Kränzler, F.~Brand, K.~Fischer, D.~T. Nguyen, A.~Regensky, and A.~Kaup, ``The bjøntegaard bible why your way of comparing video codecs may be wrong,'' \emph{IEEE Transactions on Image Processing}, vol.~33, pp. 987--1001, 2024.

\bibitem{sun2017wspsnr}
Y.~Sun, A.~Lu, and L.~Yu, ``Weighted-to-spherically-uniform quality evaluation for omnidirectional video,'' \emph{IEEE Signal Processing Letters}, vol.~24, no.~9, pp. 1408--1412, 2017.

\bibitem{balle2016}
J.~Ball\'e, V.~Laparra, and E.~P. Simoncelli, ``End-to-end optimization of nonlinear transform codes for perceptual quality,'' in \emph{Proc. Picture Coding Symposium (PCS)}, 2016, pp. 1--5.

\bibitem{Minnen2020channel-cm}
D.~Minnen and S.~Singh, ``Channel-wise autoregressive entropy models for learned image compression,'' in \emph{Proc. IEEE International Conference on Image Processing (ICIP)}, 2020, pp. 3339--3343.

\bibitem{Liang2021channel-grouped-cm}
L.~Yuan, J.~Luo, S.~Li, W.~Dai, C.~Li, J.~Zou, and H.~Xiong, ``Learned image compression with channel-wise grouped context modeling,'' in \emph{Proc. IEEE International Conference on Image Processing (ICIP)}, 2021, pp. 2099--2103.

\bibitem{xiao2012sun360}
J.~Xiao, K.~A. Ehinger, A.~Oliva, and A.~Torralba, ``Recognizing scene viewpoint using panoramic place representation,'' in \emph{Proc. IEEE/CVF Conference on Computer Vision and Pattern Recognition (CVPR)}, 2012, pp. 2695--2702.

\bibitem{yu2015spsnr}
M.~Yu, H.~Lakshman, and B.~Girod, ``A framework to evaluate omnidirectional video coding schemes,'' in \emph{Proc. IEEE International Symposium on Mixed and Augmented Reality (ISMAR)}, 2015, pp. 31--36.

\end{thebibliography}

\end{document}